\documentclass[aps,pre,reprint,longbibliography,floatfix]{revtex4-1}

\usepackage{amsmath}
\usepackage{amssymb}
\usepackage{graphicx,psfrag}
\usepackage{verbatim}
\usepackage{xcolor}
\usepackage{ulem}
\usepackage [english]{babel}
\usepackage [autostyle, english = american]{csquotes}
\MakeOuterQuote{"}
\usepackage{appendix}
\usepackage{siunitx}
\usepackage{mathtools}
\usepackage{subcaption}

\newcommand\feqi{f^\text{eq}_i}

\begin{document}

\title{A Method for Analytical Solutions in the Lattice Boltzmann Method}


\author{Jordan Larson}
\email[]{larso169@purdue.edu}
\affiliation{Department of Mathematics, Purdue University, West Lafayette, Indiana 47907, USA}
\author{Alexander J. Wagner}
\email[]{alexander.wagner@ndsu.edu}
\homepage[]{www.ndsu.edu/pubweb/$\sim$carswagn}

\affiliation{Department of Physics, North Dakota State University, Fargo, North Dakota 58108, USA}


\date{\today}

\begin{abstract}
Analytical solutions to the lattice Boltzmann Equation make it possible to study the method itself, explore the properties of its collision operator, and identify implementations of boundary conditions. In this paper, we propose a method to find analytical solutions where the macroscopic flow profile is known. We test this method on bulk Couette flow aligned and inclined to the simulation lattice with the quadratic and entropic equilibrium distributions. Our method indeed provides an analytical solution to these flows when using the quadratic distribution. When the flow is aligned to the lattice, our method provides an analytical solution using the entropic distribution for practical relaxation times and shear rates. We show that a small even order truncation of the formal solution is optimal for accuracy-compute-time trade-off. In the inclined case, our method does not conserve momentum, by a small relative error, when using the entropic distribution. We also discover that entropic lattice Boltzmann method is not compatible with the angled Couette flow. We discuss the application of our method to more complicated flows.
\end{abstract}
\maketitle

\section{Introduction}
Lattice Boltzmann methods have emerged as a powerful tool to simulate fluids. These methods were originally based on averaging lattice gas methods and therefore unconditionally stable \cite{higuera1989simulating}. Lattice Boltzmann methods became even more versatile by breaking the link to lattice gases by allowing for arbitrary local equilibrium distributions \cite{higuera1989boltzmann,qian1992lattice}. One advantage of this method is the relative ease with which complex geometries can be implemented, although sub-lattice resolution for boundary conditions remain an area of current research \cite{Ginzburg2023}.

We define an analytical solution to be a solution of both the lattice Boltzmann Equation and the Navier-Stokes Equations. It is highly desirable to find such solutions to known flow profiles. Such knowledge would prove the lattice Boltzmann method recovers the analytical solutions directly or would guide development on novel lattice Boltzmann methods that can.

Analytical solutions have previously been derived for some specific flow profiles. Zou \textit{et al.} \cite{zou1995analytical} were the first to derive analytical solutions for Poiseuille and Couette flows aligned to the lattice using a quadratic equilibrium distribution and BGK collision operator. We show that our solution applied to the Couette flow recovers their result (and we identify a misprint in \cite{zou1995analytical}).

Ginzburg \textit{et al.} \cite{Ginzburg2023} analyzed Couette flow (particularly in the more complex version including a normal injection leading to an exponential velocity profile) and found analytical solutions for an inclined Couette flow in a two relaxation time scheme. This analysis shows the utility of analytical solutions to guide the development of better boundary conditions. The derived solutions, however, are only exact for some special "magic values" for the relaxation times and the specific equilibrium distribution considered. Our approach is somewhat different. Here we present a formal analytical solutions for an arbitrary equilibrium distribution but only for a single relaxation time scheme, and only for flows for which the flow solution is know.

This is similar in spirit to the work of H\'azi \cite{Hazi2003}. They presented a method to check the consistency of the lattice Boltzmann equation with a known analytical flow description, but only for full relaxation to local equilibrium. The basis for this idea is that local density and velocity completely define the local equilibrium distribution. Somewhat surprisingly, H\'azi did not recover the known analytical solutions but instead used his results to analyze presumed error terms. These discrepancies were due to the fact that he used non-standard forcing terms for the lattice Boltzmann method, leading to the errors he observed. In contrast, our analysis is valid for arbitrary relaxation times.

Ansumali \textit{et al.} \cite{Ansumali2007} also highlighted the importance of analytical solutions by considering a time and space continuous lattice Boltzmann formulation. They found analytical solutions for non-inclined Couette flows in arbitrary Knudsen numbers. They showed that a next nearest neighbor approximation was insufficient to fully recover boundary conditions at high Knudsen numbers. 

Johnson \textit{et al.} \cite{Johnsonetal2024} found an analytical solution for the bounded Couette flow using special boundary conditions of Maxwell-type. They compared their solution to molecular dynamics simulations and found excellent agreement. \cite{Johnsonetal2024} exhibits the utility of analytical solution to validate lattice Boltzmann methods. 

Although the reviewed literature used the popular quadratic equilibrium distribution to derive their analytical solutions, the entropic equilibrium distribution is another important choice. This is derived by minimizing an H-functional \cite{ansumali2003minimal} and also results from the lattice Boltzmann limit of an integer lattice \cite{blommel2018integer}. Entropic lattice Boltzmann has multiple advantages, as illustrated by Frapolli \textit{et al.} \cite{Frapolli2020}, such as its ability to handle higher mach-number flows and its greater stability. Thus, in addition to considering the standard quadratic equilibrium distribution, we extend the analytical solutions to entropic lattice Boltzmann.

We identify a formal analytical solution of the lattice Boltzmann method when the macroscopic flow fields are known. We then examine this formal solution to determine its consistency with the lattice Boltzmann method. The first application of the new approach is on a Couette flow that can be rotated with respect to the lattice by an arbitrary angle. In this case, we show that the analytical solution for a quadratic equilibrium distribution is consistent with the lattice Boltzmann Equation. However, the solution for the entropic distribution is more complicated. Machine-accuracy agreement is observed for practical relaxation times and shear rates when the Couette flow is aligned with the lattice. When the flow is inclined, we observe deviations between the analytical flow profile and the entropic lattice Boltzmann method.

This paper observes the following outline. First, we introduce the lattice Boltzmann method. Second, we explain the formal the formal analytical solution. Third, we introduce the aligned Couette flow, and we apply the formal analytical solution to this problem for both the polynomial and entropic local equilibrium distributions. Fourth, we extend the analysis to the inclined Couette flow. Finally, we conclude by discussing potential applications of this approach to more complex flow situations. 

\section{Lattice Boltzmann Method}
In the lattice Boltzmann method, probability densities $f_i(\mathbf{x}, t)$, $i \in I$ for some finite index set $I$, are assigned to each lattice node $\mathbf{x}$ at time $t$ which have velocity vector $\mathbf{v}_i$. We require that $\mathbf{x} + \mathbf{v}_i \Delta t$ to be another node. When lattice Boltzmann methods are derived from underlying lattice gas methods, the $f_i$ represent the expectation value for particle occupation numbers for particles moving from lattice cell $x-v_i$ at time $t-\Delta t$ to lattice site $x$ at time $t$. The $f_i(x,t)$ densities evolve according to the lattice Boltzmann Equation \cite{KrugerLatticeBoltzmann}:
\begin{equation}
    f_i(\mathbf{x} + \mathbf{v}_i \Delta t, t + \Delta t) = f_i(\mathbf{x}, t) + \Omega_i,
    \label{eqn: Lattice Boltzmann Equation with arbitrary col operator}
\end{equation}
where $\Omega_i$ is the collision operator. The simplest collision operator in the lattice Boltzmann method, proposed by \cite{qian1992lattice}, is
\begin{equation}
    \Omega_i = \frac{\Delta t}{\tau} \left[\feqi(\rho, \mathbf{u}) - f_i(\mathbf{x}, t) \right].
    \label{BGKGeneral}
\end{equation}
Collisions bring the $f_i$ densities closer to a local equilibrium distribution $f_i^\text{eq}$, which only depends on the locally conserved $\rho$ and $\mathbf{u}$. Relaxation time $\tau$ controls the rate of convergence. To ensure conservation of mass and momentum, we have the following moment equations:
\begin{align}
        \rho(\mathbf{x},t) &= \sum_i f_i(\mathbf{x},t), \label{eqn:rhomoment} \\
        \rho \mathbf{u}(\mathbf{x}, t) &= \sum_i f_i(\mathbf{x},t) \mathbf{v}_i.
        \label{eqn:momentummoment}
    \end{align}
$f_i^\text{eq}$ must also satisfy these moments. Energy in general is not conserved, and temperature is forced to be constant through the local equilibrium distribution.

To simplify notation, we adopt lattice units $\Delta t = 1$, where $\Delta t$ is the time advanced after each iteration, and $\Delta x = \min\{||\mathbf{v}_i \Delta t|| : i \in I\} = 1$.

\section{Analytical Solution}
To derive our formal analytical solution we start with the lattice Boltzmann Equation (\ref{eqn: Lattice Boltzmann Equation with arbitrary col operator}) with collision operator (\ref{BGKGeneral}):
\begin{equation}
    f_i(\mathbf{x} + \mathbf{v}_i, t + 1) = f_i(\mathbf{x}, t) + \frac{1}{\tau}[\feqi(\rho, \mathbf{u}) - f_i(\mathbf{x}, t)].
    \label{eqn: BGKLB}
\end{equation}
As is usual, we assume that the discrete $f_i$ has an extension to a smooth function and expand the left hand side of equation (\ref{eqn: BGKLB}) via a Taylor series centered at $(\mathbf{x}, t)$ to find
\begin{equation}
    f_i(\mathbf{x} + \mathbf{v}_i, t + 1) = \sum_{n=0}^\infty \frac{1}{n!} \left( \partial_t + v_{i \alpha} \partial_\alpha\right)^n f_i(\mathbf{x},t),
    \label{eqn: Taylor}
\end{equation}
with Einstein notation used. Note that this equality is formal since \textit{a priori} the infinite series may not even converge. We insert this into equation (\ref{eqn: BGKLB}) and get
\begin{equation}
    f_i(\mathbf{x}, t) = \feqi (\rho, \mathbf{u})- \tau \sum_{n=1}^\infty \frac{1}{n!} \left( \partial_t + v_{i \alpha} \partial_\alpha\right)^n f_i(\mathbf{x},t).
    \label{eqn: derivation just before repeated insertion}
\end{equation}
This gives us an expression of the $f_i$ in terms of the equilibrium distribution and derivatives of the $f_i$. For more complicated collision terms $\Omega_i$, like that of the multi-relaxation time procedure, a similar approach is possible. However, the algebra is more tedious \cite{kaehler2013derivation}.

The customary trick employed at this stage is to insert equation (\ref{eqn: derivation just before repeated insertion}) into itself. Typically, this step is only done once, since the Navier-Stokes equations have at most second-order derivatives. Instead, we continue the process to fully eliminate the $f_i$ on the right-hand side of the equation:
\begin{align}
    f_i(\mathbf{x}, t) =& f_i^\text{eq} \nonumber \\
    &- \tau \sum_{n=1}^\infty \frac{1}{n!}\left(\partial_t + v_{i \alpha} \partial_\alpha \right)^n f_i^{\text{eq}} \nonumber \\
    &+ \tau^2 \sum_{n=1}^\infty \sum_{m=1}^\infty \frac{1}{n!m!} \left(\partial_t + v_{i \alpha } \partial_\alpha \right)^{n+m} f_i^\text{eq} \nonumber \\
    &\mp \cdots .
    \label{eqn: Fully Expanded fi just after derivation}
\end{align}
Equation (\ref{eqn: Fully Expanded fi just after derivation}) gives $f_i(\mathbf{x},t)$ solely in terms of the macroscopic quantities $\rho$ and $u$. Collecting all terms with the same powers of the derivatives, we write equation (\ref{eqn: Fully Expanded fi just after derivation}) as 
\begin{equation}
    f_i(\mathbf{x}, t) = \sum_{n = 0}^\infty P_n(\tau) (\partial_t + v_{i\alpha}\partial_\alpha)^n \feqi (\rho, \mathbf{u}),
    \label{eqn: simplified fi completely analytical general solution with polynomials}
\end{equation}
where the $P_n(\tau)$ are polynomials that satisfy the recurrence relation
\begin{align}
    P_0(\tau) &= 1, \\
    P_n(\tau) &= -\sum_{k = 1}^n\frac{\tau}{k!} P_{n - k}(\tau).
    \label{eqn: P_n polynomials}
\end{align}
A. J. Wagner previously derived this result up to $n=4$ in \cite{wagner2006thermodynamic}. We use chain rule to evaluate the derivatives of the equilibrium distribution in equation (\ref{eqn: simplified fi completely analytical general solution with polynomials}):
\begin{align*}
    \partial_t \feqi &= \frac{\partial \feqi}{\partial \rho} \partial_t \rho + \frac{\partial \feqi}{\partial u_\beta} \partial_t u_\beta,\\
    \partial_\alpha \feqi &= \frac{\partial \feqi}{\partial \rho} \partial_\alpha \rho+ \frac{\partial \feqi}{\partial u_\beta} \partial_\alpha u_\beta.
\end{align*}
We give the fully expanded equation for the two-dimensional case in Appendix \ref{appenddix: big equations} as equation (\ref{eqn: huge chain rule with polynomial}) because it is too lengthy to present here.

This gives a formal representation for fully analytical solution for the $f_i(\mathbf{x},t)$ for situations where the analytical solution for the hydrodynamic fields $\rho(\mathbf{x},t)$ and $u(\mathbf{x},t)$ is known. There are several caveats. First, the $f_i(\mathbf{x},t)$ are only defined on discrete lattice points, so the exact meaning of the continuous derivatives in Eq. (\ref{eqn: Taylor}) is somewhat suspect. Second, the procedure of inserting equation (\ref{eqn: derivation just before repeated insertion}) into itself is not guaranteed to converge. These steps are commonly used in the analysis of lattice Boltzmann methods, and the ensuing analysis here facilitates evaluation of the appropriateness of these operations. Third, a necessary condition for our method to give an analytical solution is that $\rho(\mathbf{x},t)$ and $\mathbf{u}(\mathbf{x},t)$ must correspond to solutions of the lattice Boltzmann equation (\ref{eqn: BGKLB}).

Ideally, to validate the solution, we would insert equation (\ref{eqn: simplified fi completely analytical general solution with polynomials}) into the lattice Boltzmann equation (\ref{eqn: BGKLB}) and check consistency. We do not know how to analyze this in the most general case due to the presence of discrete space and time terms along with infinite orders of derivatives of $\rho$ and $\mathbf{u}$. However, the formal solution simplifies greatly for the Couette flow using the quadratic equilibrium distribution, and we can directly verify consistency. The formal solution does not sufficiently simplify in the case of entropic lattice Boltzmann, so we evaluate the result numerically.

\section{Numerical Examination}
The analytical solution was derived under the assumption that the discrete $f_i(\mathbf{x},t)$ extends to an infinitely-differentiable function. This section examines how well this assumption agrees with the results of numerical lattice Boltzmann simulations.

In order to test the solution, we choose a specific velocity set $\{\mathbf{v}_i\}$ and a specific equilibrium distribution $f_i^\text{eq}$. We use the following two-dimensional velocity set with lattice units: 
\begin{equation}
    \mathbf{v}_i = \begin{cases}
        \begin{pmatrix}
                0\\
                0
            \end{pmatrix} & i = 0,\\
        \begin{pmatrix}
            \cos \left(\frac{\pi[i-1]}{2} \right)\\
            \sin \left( \frac{\pi[i-1]}{2} \right)
        \end{pmatrix} & i = 1,2,3,4,\\
        \begin{pmatrix}
            \sqrt{2}\cos \left(\frac{\pi[i-9/2]}{2} \right)\\
            \sqrt{2}\sin \left( \frac{\pi[i-9/2]}{2} \right)
        \end{pmatrix} & i = 5,6,7,8,
    \end{cases}
\end{equation}
often referred to as D2Q9. The quadratic local equilibrium distribution, first proposed by Qian \cite{qian1992lattice}, is
\begin{align}
&f_i^{\text{eq, pol}}(\rho, \mathbf{u})\nonumber\\
=& \rho w_i \left(1 + \frac{1}{c_s^2} u_\alpha v_{i\alpha} + \frac{1}{2c_s^4} (u_\alpha v_{i\alpha})^2 
- \frac{1}{2c_s^2} u_\alpha u_\alpha \right),
\label{quadGeneral}
\end{align}
with Einstein notation used. The weights $w_i$ are a product of one-dimensional weights for velocities $v_{i\alpha}$:
\begin{equation}
  w_{v_{i\alpha}} = \frac{2}{3} - \frac{1}{2}v_{i\alpha}^2,
  \label{eqn: w_via formula}
\end{equation}
and we put $w_i = \prod_\alpha w_{v_{i\alpha}}$. Also, $c_s$ is the isothermal sound speed, which satisfies $c_s^2 = \frac{1}{3}$ in lattice units. 

Developed by Ansumali \textit{et al.} \cite{Ansumali2007} as minimizing an $H$-functional $H=\sum_i f_i \ln(f_i/w_i)$, the entropic equilibrium distribution is the only known equilibrium distribution derivable from an underlying lattice gas description as shown by Blommel \textit{et al.} \cite{blommel2018integer}. It is
\begin{align}
    &f_i^\text{eq,ent}(\rho, \mathbf{u}) \nonumber\\
    =& \rho \prod_{\alpha} w_{v_{i\alpha}} \left[1 + \frac{v_{i\alpha} u_\alpha}{c_s^2} + \left(\frac{v_{i\alpha}^2}{c_s^2} - 1\right) \left(\sqrt{1 + \frac{u_\alpha^2}{c_s^2}} - 1 \right) \right],
    \label{eqn:entropicDistribution}
\end{align}
with Einstein notation not used in this formula.

We now consider the Couette flow aligned to the D2Q9 lattice in order to analyze our method. Later, we will consider Couette flow inclined by an arbitrary angle with respect to the lattice.

\subsection{Aligned Couette Flow}
\label{sec: Numerical Examination of Aligned Shear Flow}
We test the solution on one of the simplest solutions to the Navier-Stokes equations. The hydrodynamic fields for a fluid in Couette flow aligned with the $x$-direction are

\begin{align}
    \rho(\mathbf{x},t) &= 1 \label{eqn: Flat Couette Rho Profile},\\
    \mathbf{u}(\mathbf{x},t) &=\left(\begin{array}{c} \dot{\gamma} y\\ 0 \end{array}\right).
    \label{eqn: Flat Couette Velocity Profile}
\end{align}
We choose units in such a way to obtain a constant density of $1$, and $\dot{\gamma}$ is some choice of shear rate in lattice units. We bound the domain of simulation by $-H \leq y \leq H$. Due to the $x$-translational symmetry of the Couette flow, we do not need to bound the $x$ dimension in our simulations, and we may use periodic boundary  conditions. This flow profile has the property that the first derivatives $\partial_y u_x$ are constant and all higher derivatives vanish and all time derivatives vanish, which considerably simplifies the proposed solution (\ref{eqn: simplified fi completely analytical general solution with polynomials}). 

In the following, let $f_i^\text{th}$ denote one of the formal analytical solutions given by equation (\ref{eqn: simplified fi completely analytical general solution with polynomials}) with the appropriate simplifications derived below. Note that in the aligned case $f_i^\text{th}$ is a function of $y$ alone.

In an experimental system, we typically achieve Couette flow by driving the fluid out of equilibrium using parallel walls moving a constant relative velocity along the parallel direction. We must configure the following simulations with initial and boundary conditions to induce the Couette flow. The initial conditions are inconsequential, as explained below. For the boundary conditions, we will inject certain probability densities on the boundary sites $\mathbf{x}_b$. Because the above derivation ignored boundary conditions, we expect the proposed solution to only hold in the bulk of the final simulation result due to boundary effects. The simplest approach to nullify boundary effects and facilitate the use of smaller simulation sizes is to put
\begin{align}
    f_i(x,-H,t) &= f_i^\text{th}(-H), \label{eqn: Aligned boundary condition left}\\
    f_i(x,H,t) &= f_i^\text{th}(H). \label{eqn: Aligned boundary condition right}
\end{align}
It is necessary that these densities $f_i^\text{th}$ are analytical solutions to the Couette flow profile and consequently their moments induce the Couette flow profile. Equations (\ref{eqn: Aligned boundary condition left}) and (\ref{eqn: Aligned boundary condition right}) are formal boundary conditions because we have not checked that $f_i^\text{th}$ from equation (\ref{eqn: simplified fi completely analytical general solution with polynomials}) is in the most general case an analytical solution. Even if the proposed $f_i^\text{th}$ is not a solution and has incorrect moments, we may still use this method to numerically evaluate whether this works as a solution.

We make some simulation choices to speed up calculations. First, because the Couette flow is steady-state, the initial condition is unimportant. Thus, to speed up convergence in some cases, we set the initial condition to be $f_i(x, y, 0) = f_i^\text{th}(y)$. Second, we are justified to take $H = 3$ lattice units in the ensuing tests of the aligned flow because if $f_i^\text{th}$ is indeed an analytical solution, there would be no boundary effects. Third, the initial condition allows us to use only one time step in the simulations because $f_i(y, 1) = f_i^\text{th}(y)$ if and only if $f_i^\text{th}$ is an analytical solution.

It appears that using a different function in the insertion boundary conditions, which has the correct moments, does not induce the expected bulk velocity $\mathbf{u}(y)$ or density profiles $\rho$ of the Couette flow in the steady-state limit of iterations in general. Inserting any order (including the equilibrium distribution itself) of the solution corresponding to the entropic distribution gives macroscopic agreement for all iterations at $\tau = 1$ only. The same applies to orders $0$ and $1$ of the solution associated with the quadratic distribution. Furthermore, we noticed that these simulations converge to a unique bulk solution given a large enough simulation. Consequently, this hints at the uniqueness of the solution given by equation (\ref{eqn: simplified fi completely analytical general solution with polynomials}) as applied to the aligned Couette flow. Further investigation is needed for this claim.  

Another way to implement invisible boundaries for a Couette flow that avoids physical effects like wall slip or a Knudsen layer is to use Lees-Edwards boundary conditions \cite{wagner2002lees}. This method applies a Galilean transformation to the $f_i$ densities at the top and bottom boundaries to induce the velocity shear. The existence of an analytical solution whose injection on the boundaries causes a similar shear profile can assist with Lees-Edwards implementation. If such a solution does not exist, as in the case of inclined Couette flow in entropic lattice Boltzmann, then boundary effects are unavoidable. 

\subsubsection{Polynomial Equilibrium}
Using the quadratic equilibrium distribution, velocity derivatives of $f_i^\text{eq, pol}$ of order greater than $2$ vanish because equation (\ref{quadGeneral}) is quadratic in $u_\alpha$. Equation (\ref{eqn: simplified fi completely analytical general solution with polynomials}) significantly simplifies, and we obtain the following solution:
\begin{align}
    f_i^\text{th, pol}(y) =& w_i \rho \left(1 + 3v_{ix}u_x + \frac{3}{2}u_x^2[3v_{ix}^2 - 1] \right) \nonumber \\
    &-3\tau w_i \rho v_{iy} \dot{\gamma} (v_{ix} + u_x[3v_{ix}^2 - 1]) \nonumber \\
    &+3\tau(\tau - \frac{1}{2})w_i \rho v_{iy}^2 \dot{\gamma}^2 (3v_{ix}^2-1).
    \label{eqn: quadratic couette flat analytical solution}
\end{align}

We may prove this to be an analytical solution by inserting it into equation (\ref{eqn: BGKLB}). Rewrite (\ref{eqn: BGKLB}) as
\begin{equation}
    \tau[f_i^\text{th, pol}(y+v_{iy}) - f_i^\text{th, pol}(y)] + f_i^\text{th, pol}(y) - f_i^\text{eq, pol}(y)=0.
\end{equation}
We may simplify these terms as the following:
\begin{align}
    &f_i^\text{th, pol}(y+v_{iy}) - f_i^\text{th, pol}(y)\nonumber\\ =& w_i \rho \left(3v_{ix}v_{iy}\dot{\gamma} + \frac{3}{2}\dot{\gamma}^2v_{iy}(2y + v_{iy})(3v_{ix}^2-1) \right. \nonumber\\
    &- \left. 3\tau v_{iy}^2\dot{\gamma}^2(3v_{ix}^2 - 1)\right),
\end{align}
and
\begin{align}
    &f_i^\text{th, pol}(y) - f_i^\text{eq, pol}(\rho, u_x) \nonumber\\
    =& -3\tau w_i \rho v_{iy} \dot{\gamma} [v_{ix} + u_x(3v_{ix}^2 - 1)] \nonumber \\
    &+3\tau(\tau - \frac{1}{2})w_i \rho v_{iy}^2 \dot{\gamma}^2 (3v_{ix}^2-1).
\end{align}
Inserting these results, equation (\ref{eqn: quadratic couette flat analytical solution}) indeed fulfills equation (\ref{eqn: BGKLB}).

Zou \text{et al.} \cite{zou1995analytical} previously derived this solution (barring a typo in their $f_7$ formula and a reference frame shift) using a different approach.

Although we have demonstrated that equation (\ref{eqn: quadratic couette flat analytical solution}) is indeed an analytical solution, we cannot do so for the analogous proposal for the entropic distribution and must solely rely on numerical analysis. As a sort of control feature, we will run numerical tests on equation (\ref{eqn: quadratic couette flat analytical solution}). To numerically compare simulation to prediction, we define the error at a certain $y$ position by
\begin{equation}
    (\Delta f)(y) = \sqrt{\frac{1}{9} \sum_{i = 0}^8 \left[ \frac{f_i^{\text{sim}}(y) - f_i^{\text{th}}(y)}{f_i^\text{eq}(\rho(y), \mathbf{u}(y) )} \right]^2},
    \label{eqn: 1D error measure}
\end{equation} where $f_i^\text{sim}$ refers to the population data given by running a lattice Boltzmann simulation with associated parameter levels for a certain number of iterations; $f_i^\text{th}$ is the analytical solution being tested; $f_i^\text{eq}$ is the appropriate equilibrium distribution; and $\rho$ and $\mathbf{u}$ are given by equations (\ref{eqn: Flat Couette Rho Profile}) and (\ref{eqn: Flat Couette Velocity Profile}).

\begin{figure}
    \centering
    \includegraphics[width=\linewidth]{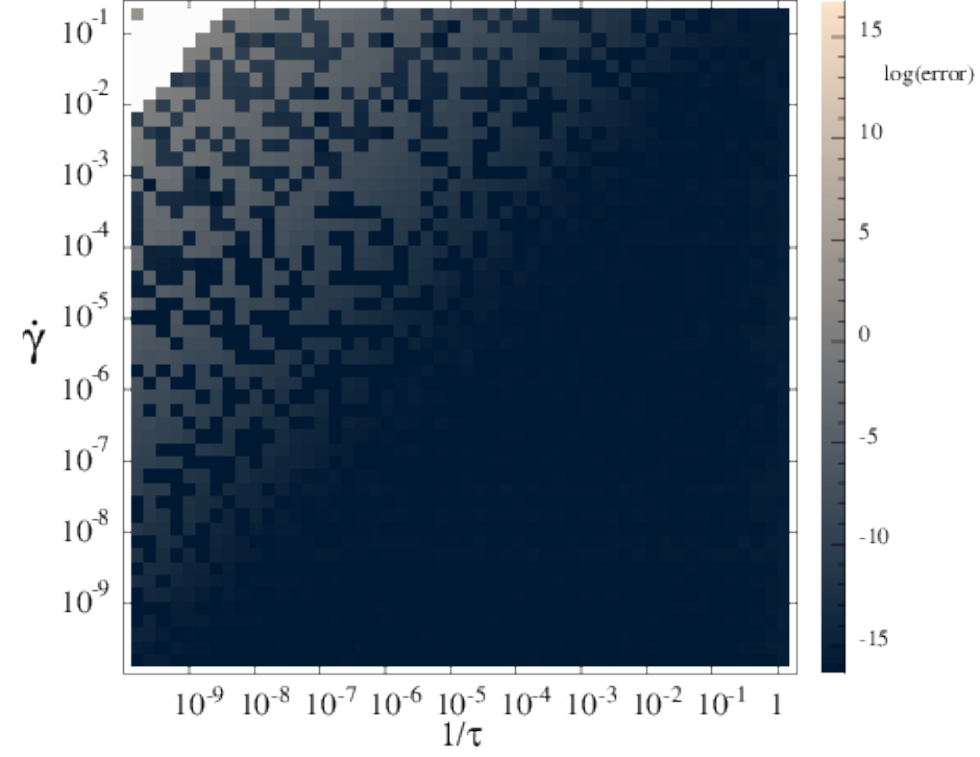}
    \caption{Heat map of $\log_{10}((\Delta f)(y_0))$ where $(\Delta f)(y_0)$ is given by equation (\ref{eqn: 1D error measure}). Equation (\ref{eqn: quadratic couette flat analytical solution}) defines $f_i^\text{th, pol}$ and we fix $y_0 = 0$. To generate the error at a certain point, we run a simulation for $1$ time step with the quadratic equilibrium distribution and the given $\tau$, $\dot{\gamma}$, $y_0$, along with the other parameters mentioned at the beginning of section \ref{sec: Numerical Examination of Aligned Shear Flow}. The choice of $y_0 = 0$ is not too important, but it does simplify formulae to some degree since $u_x = 0$. This heat map and the following graphs depend on the specific $y_0$ chosen. There are $50$x$50 = 2500$ data points.}
    \label{fig: 1D Quadratic Heat Map}
\end{figure}

Figure \ref{fig: 1D Quadratic Heat Map} shows the results of simulations of one iteration testing $f_i^\text{th, pol}$ of equation (\ref{eqn: quadratic couette flat analytical solution}) on various combinations of $\dot{\gamma}$ and $\frac{1}{\tau}$.

We observe non-machine accuracy error in Figure \ref{fig: 1D Quadratic Heat Map} despite the proof that equation (\ref{eqn: quadratic couette flat analytical solution}) is an analytical solution to the lattice Boltzmann equation (\ref{eqn: BGKLB}). We surmise this arises from round-off error since we use doubles in C, which can accurately store at most 15 significant figures, to generate the data in the Figure. Near the top left corner of Figure \ref{fig: 1D Quadratic Heat Map}, $f_i^\text{th, pol}$ is on the order of $10^{20}$, and we inject it on the boundary to run the corresponding simulation. Large errors thereby propagate through the simulation and in some cases $f_i^\text{sim}$ returns as "nan," indicated by white spots in the Figure. Interestingly, the diagonal line given by $\tau = \frac{1}{\dot{\gamma}}$ clearly separates the machine-accuracy region from the erroneous region, which is expected because the leading order of (\ref{eqn: quadratic couette flat analytical solution}) is in terms of $(\tau \dot{\gamma})^2$.

In summary, despite its analyticity, there is a practical range for valid use of the solution due to numerical limitations. These limitations present a confounding variable in the analysis of the entropic solution, for which we only consult numerical evidence of convergence. 

\subsubsection{Entropic Equilibrium}
\label{sec: numerical examination, aligned entropic case}
The choice of equilibrium distribution has an essential effect on the simplification of equation (\ref{eqn: simplified fi completely analytical general solution with polynomials}). Because the entropic equation uses square roots, the derivatives of $f_i^\text{eq, ent}$ do not vanish and we require a formal infinite series. Equation (\ref{eqn:FlatCouetteEntropicAnalSol}) is the proposed analytical solution using the entropic equilibrium distribution, which we give in the appendix because it is too lengthy to write here. Since equation (\ref{eqn:FlatCouetteEntropicAnalSol}) has not been checked to satisfy equation (\ref{eqn: BGKLB}), we do not know if it is indeed an analytical solution. However, in most practical situations of reasonable $\tau$ and $\dot{\gamma}$ values, low order truncations have small error in simulating a Couette flow.

Due to numerical limitations, we test truncated versions of the proposed solution. We define
\begin{equation}
    (f_i^\text{th, ent})_N \coloneq \sum_{n = 0}^N f_i^\text{th, ent, n},
    \label{eqn: 1D Entropic Solution Truncation}
\end{equation}
where we define $f_i^\text{th, ent, n}$ by equation (\ref{eqn: fithentN}). We will refer to $N$ as the "order truncation" or "order" being tested.

\begin{figure}
    \centering
    \includegraphics[width=\linewidth]{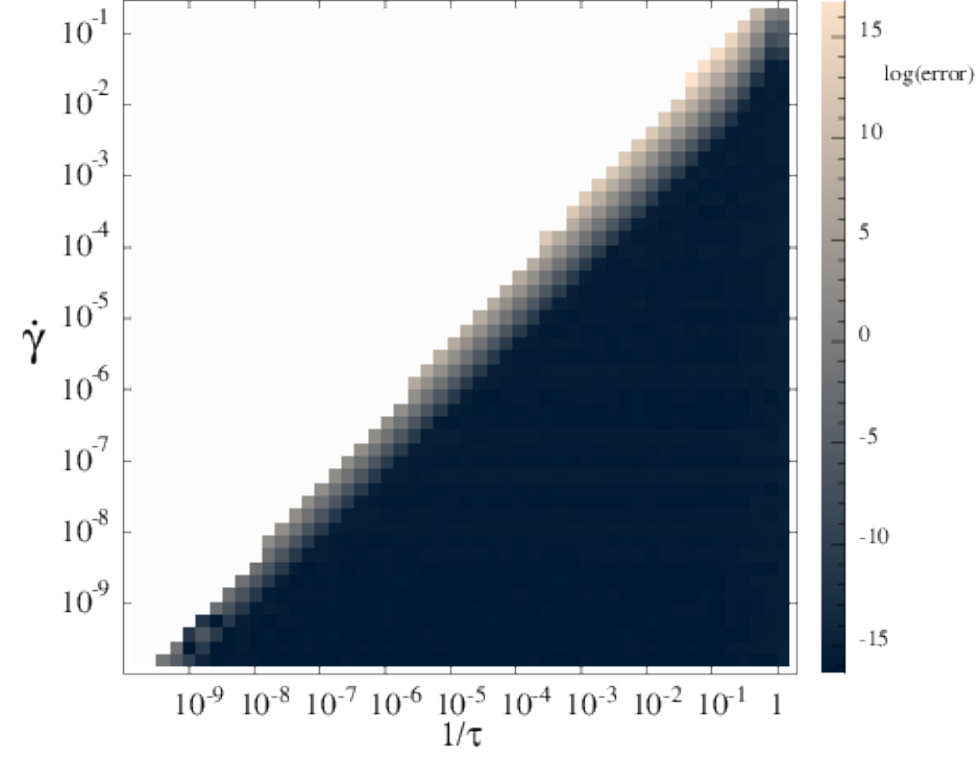}
    \caption{Heat map of $\log_{10}((\Delta f)(y_0))$ of equation (\ref{eqn: 1D Entropic Solution Truncation}), with $y_0 = 0$ fixed and $N = 22$. The data generation procedure is the same as that in Figure \ref{fig: 1D Quadratic Heat Map}.}
    \label{fig: 1D Entropic Heat Map}
\end{figure}

Figure \ref{fig: 1D Entropic Heat Map} shows the results of simulations of one iteration testing $(f_i^\text{th, ent})_{22}$ exhausting combinations of $\dot{\gamma}$ and $\frac{1}{\tau}$. We do not know if the non-machine-accuracy error is due to floating point error or the non-convergence of equation (\ref{eqn:FlatCouetteEntropicAnalSol}). It is possible that the series is asymptotic and higher order terms are not small, which prevents convergence. On the other hand, widespread numerical error (nan-white space) above $\tau = \frac{1}{\dot{\gamma}}$ is expected as the leading order is in terms of $(\tau \dot{\gamma})^{22}$. However, the error starts diverging for $\tau, \dot{\gamma}$ values below this line. This is due to the presence of the leading order pre-factor on the order of $10^{14}$ of equation (\ref{eqn: 1D Entropic Solution Truncation}). Despite this, error is on machine-accuracy level for the practical values of $\tau, \dot{\gamma}$ in widespread use.

\begin{figure}
    \centering
    \includegraphics[width=\linewidth]{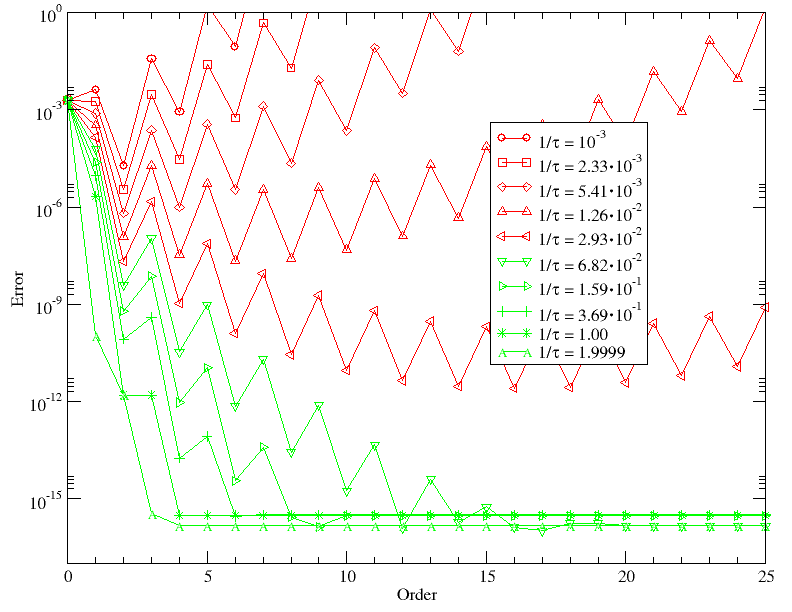}
    \caption{Plot of $(\Delta f)(y_0)$ of $(f_i^\text{th, ent})_N$ against $N$, the order of truncation. $\dot{\gamma} = 10^{-3}$ and $y_0 = 0$ are fixed for all simulations, which are run for $1$ iteration.}
    \label{fig: Entropic Error vs. Order Gdot Fixed 0.001}
\end{figure}

\begin{figure}
    \centering
    \includegraphics[width=\linewidth]{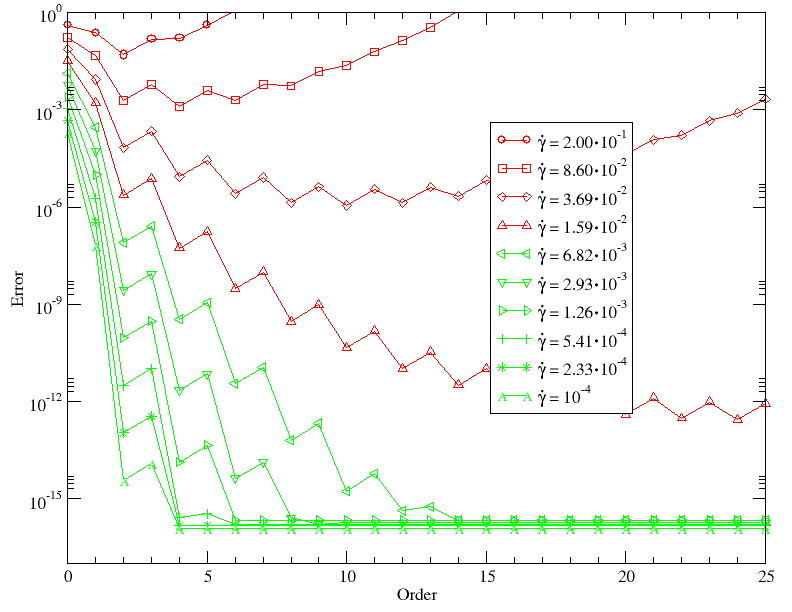}
    \caption{Plot of $(\Delta f)(y_0)$ of $(f_i^\text{th, ent})_N$ against $N$. Data generation is the same as in Figure \ref{fig: Entropic Error vs. Order Gdot Fixed 0.001} except that $\tau = 2$ is fixed and $\dot{\gamma}$ range over certain values.}
    \label{fig: Entropic Error vs. Order Omega Fixed 0.5}
\end{figure}

\begin{figure}
    \centering
    \includegraphics[width=\linewidth]{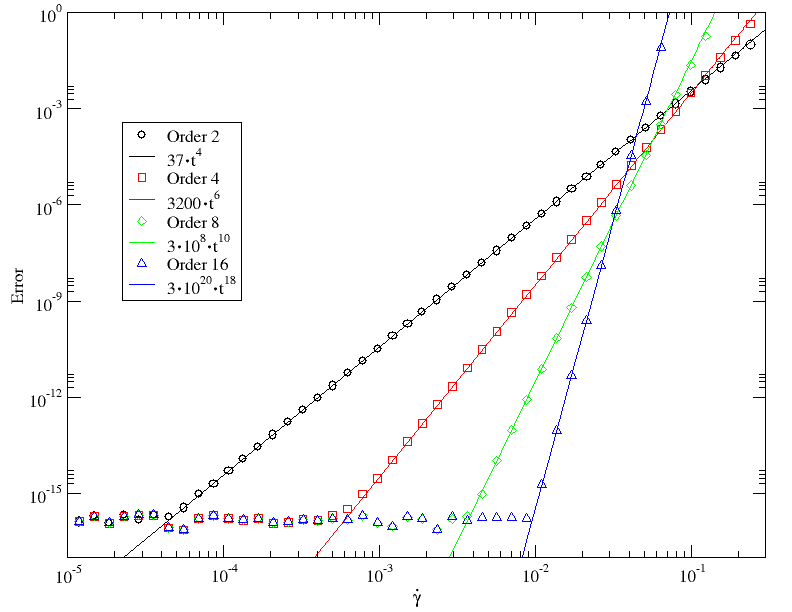}
    \caption{Convergence results for different shear rates for $\tau = 2$ for different order $N$ of $(f^\text{th, ent})_N$.}
    \label{fig: Entropic Error vs. gdot Omega 0.5}
\end{figure}

\begin{figure}
    \centering
    \includegraphics[width=\linewidth]{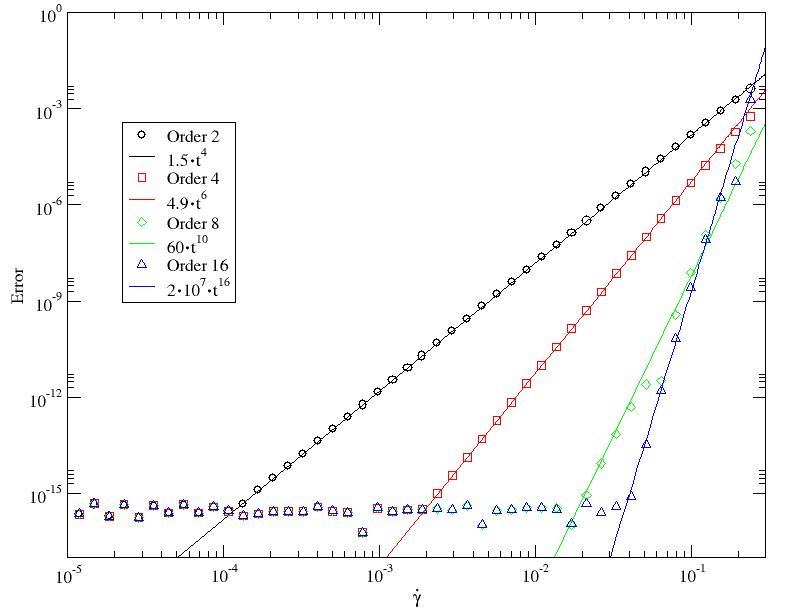}
    \caption{Convergence results for different shear rates for $\tau = \frac{2}{3}$ for different order $N$ of $(f^\text{th, ent})_N$.}
    \label{fig: Entropic Error vs. gdot Omega 1.5}
\end{figure}

Figures \ref{fig: Entropic Error vs. Order Gdot Fixed 0.001} and \ref{fig: Entropic Error vs. Order Omega Fixed 0.5} show how the error $\Delta f$ of equation (\ref{eqn: 1D Entropic Solution Truncation}) depends on the order of truncation. We note that the mass and momentum moments of $(f_i^\text{th, ent})_1$ replicate the correct macroscopic flow profile given by equations (\ref{eqn: Flat Couette Rho Profile}) and (\ref{eqn: Flat Couette Velocity Profile}) despite it not being an analytical solution. Vertical cross sections through Figure \ref{fig: Entropic Error vs. Order Gdot Fixed 0.001} indicate how the horizontal $\dot{\gamma} = 10^{-3}$ cross section in Figure \ref{fig: 1D Entropic Heat Map} changes with order truncation. Vertical cross sections through Figure \ref{fig: Entropic Error vs. Order Omega Fixed 0.5} similarly describe vertical cross sections through the heat map. Thus, we may infer that higher order improves the accuracy of the proposed solution along cross-sections of the heat map for certain $\dot{\gamma}, \tau$ values. Figures \ref{fig: Entropic Error vs. gdot Omega 0.5} and \ref{fig: Entropic Error vs. gdot Omega 1.5} corroborate this, which also tell of a drastic accuracy increase between orders $2$ and $8$. However, they also show an asymptotic limit to the accuracy increase as a function of order, so a small order should be used. In Figures \ref{fig: Entropic Error vs. gdot Omega 0.5} and \ref{fig: Entropic Error vs. gdot Omega 1.5}, we only plotted the orders of $2^n$ for $n = 1,2,3,4$ because these orders suffice for illustrating the limiting behavior.

There is justification for why we did not include odd orders in Figures \ref{fig: Entropic Error vs. gdot Omega 0.5} and \ref{fig: Entropic Error vs. gdot Omega 1.5}. In the vast majority of cases, odd-ordered truncations make the error worse or the same compared to the error at the previous even order according to Figures \ref{fig: Entropic Error vs. Order Gdot Fixed 0.001} and \ref{fig: Entropic Error vs. Order Omega Fixed 0.5}. Further testing indicates that this difference is maximized when $y_0$ is taken such that fluid velocity is $0$. Notably, when $u_x = 0$, $f_i^\text{th, ent, n} = 0$ if $n$ is odd, so $(f_i^\text{th, ent})_{N-1} = (f_i^\text{th, ent})_N$ if $N$ is odd. One might expect therefore that the error not change between even-odd pairs. However, in neighboring sites, the velocity is not $0$, and these will influence $f_i$ in this location. Regarding $y_0$ such that fluid velocity is not $0$, the odd orders still make the error either worse or the same. In total, odd-order truncations should never be used, and a small even order, such as $4$, has the best accuracy vs. compute-time trade-off.


We may use Figures \ref{fig: Entropic Error vs. gdot Omega 0.5} and \ref{fig: Entropic Error vs. gdot Omega 1.5} to identify the order of the proposed solution needed to maintain accuracy in certain shear regimes. The less extreme the shear rate, the lower the required order.

Figures \ref{fig: Entropic Error vs. Order Omega Fixed 0.5} and \ref{fig: Entropic Error vs. gdot Omega 0.5} both indicate that $\dot{\gamma}$ on the order of magnitude of $10^{-2}$ is the last value of $\dot{\gamma}$ that achieves machine-accuracy from the proposed analytical solution equation (\ref{eqn:FlatCouetteEntropicAnalSol}) for a fixed $\tau = 2$. The figures indicate limiting behavior toward an asymptotic $\dot{\gamma}$. The successive  Figure \ref{fig: Entropic Error vs. gdot Omega 1.5} indicates a similar finding for $\dot{\gamma}$ on the order of magnitude of $10^{-1}$ and $\tau = \frac{2}{3}$.

\subsection{Inclined Couette Flow}
\label{sec: Inclined Couette flow}
\begin{figure}
    \centering
    \includegraphics[scale=0.8]{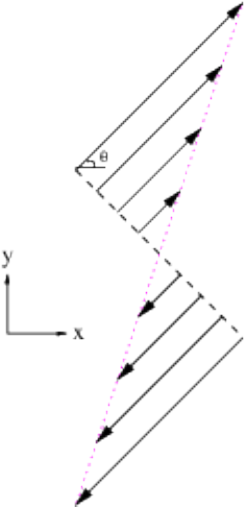}
    \caption{Inclined Couette flow. D2Q9 lattice is aligned with the standard Cartesian $xy$-coordinate grid.}
    \label{fig: angled couette flow diagram}
\end{figure}

For a fixed angle $\theta$, we describe the angled flow profile by the following:
\begin{align}
    \rho(\mathbf{x},t) &= 1 \label{eqn: Inclined Couette Rho Profile},\\
    \mathbf{u}(\mathbf{x},t) &=\left(\begin{array}{c} \dot{\gamma}[ y \cos^2(\theta) - x \sin(\theta) \cos(\theta) ] \\ \dot{\gamma}[ y \sin(\theta)\cos(\theta) - x \sin^2(\theta) ] \end{array}\right).
    \label{eqn: Inclined Couette Velocity Profile}
\end{align}
We restrict our simulations to the domain $-W \leq x \leq W$, $-H \leq y \leq H$ and take  $\theta \in [0^\circ, 360^\circ)$. See Figure \ref{fig: angled couette flow diagram} for a pictorial representation.

The boundary and initial conditions are similar to those of section \ref{sec: Numerical Examination of Aligned Shear Flow}. The initial condition is $f_i(x,y,0) = f_i^\text{th, angled}(x,y)$. For the boundary condition, if $(x,y)$ is contained on the boundary of the rectangle $[-W,W]\times[-H,H] \subseteq \mathbb{R}^2$, then
\begin{align}
    f_i(x, y) = f_i^\text{th, angled}(x,y),
\end{align}
where $f_i^\text{th, angled}$ denotes one of the formal analytical solutions derived below, depending on the one being tested at the time.

\subsubsection{Polynomial Equilibrium}
The solution proposed by equation (\ref{eqn: simplified fi completely analytical general solution with polynomials}) again truncates at order $2$ for the inclined Couette flow and quadratic local equilibrium distribution. Because of its length, we provide it in the appendix as equation (\ref{eqn: quadratic 2D Written Python}). We use a symbolic calculation feature of Python to check the consistency of the insertion of this equation into equation \ref{eqn: BGKLB}. The github repository \cite{JordanAnalSolGithub} holds the relevant files.

The python file named, "Symbolic Proof of Solving LBE of Angled Quadratic Solution.py" in \cite{JordanAnalSolGithub} provides a proof that $f_i^\text{th, angled, pol}$ is indeed an analytical solution to the inclined Couette flow.

Since the lattice Boltzmann equation (\ref{eqn: BGKLB}) is not in general rotationally symmetric, we in general cannot obtain analytical solutions by mere rotational change of variables. However, since the Couette flow and the quadratic distribution are particularly simple, it turns out this method works. More specifically, the following equality holds: 
\begin{align}
    f_i^\text{th, angled, pol}(x,y) =& f_i^\text{eq, pol}(\rho(x,y), \mathbf{u}(x,y)) \nonumber\\
    &-3\tau w_i \rho v_{i'y'} \dot{\gamma} [v_{i'x'} + u_x'(3v_{i'x'}^2 - 1)] \nonumber \\
    &+3\tau(\tau - \frac{1}{2})w_i \rho v_{i'y'}^2 \dot{\gamma}^2 (3v_{i'x'}^2-1),
    \label{eqn: factored form of angled quadratic solution}
\end{align}
where
\begin{align}
    u_x'(x,y) &= u_x\cos(\theta) + u_y \sin(\theta),\\
    v_{i'x'} &= v_{ix}\cos(\theta) + v_{iy}\sin(\theta),\\
    v_{i'y'} &= -v_{ix}\sin(\theta) + v_{iy}\cos(\theta),
\end{align}
and where equation (\ref{eqn: Inclined Couette Velocity Profile}) gives $u_x, u_y$. The reason why $f_i^\text{eq, pol}$ can take in un-rotated coordinates is because the quadratic equilibrium distribution is invariant under coordinate rotations. In particular, because it consists only of dot products, we have
\begin{equation*}
    f_{i'}^\text{eq, pol}(\rho(x',y'), \mathbf{u}(x',y')) = f_i^\text{eq, pol}(\rho(x,y), \mathbf{u}(x,y)).
\end{equation*}
The python file named, "Equivalence of the two quadratic angled expressions.py" in \cite{JordanAnalSolGithub} demonstrates the equality of  equation (\ref{eqn: factored form of angled quadratic solution}).

Certainly a necessary condition for the truth of equation (\ref{eqn: factored form of angled quadratic solution}) is that solutions for both the aligned and inclined flows exist. In general, we do not know what one may conclude about the existence of analytical solutions to the inclined version of a flow given a solution to the aligned flow and a local equilibrium distribution invariant under rotations.

\subsubsection{Entropic Equilibrium}
Once again, equation (\ref{eqn: simplified fi completely analytical general solution with polynomials}) as applied to the inclined Couette flow using the entropic local equilibrium distribution requires a (formal) infinite series. Equation (\ref{eqn: simplified fi completely analytical general solution with polynomials}) gives the formal solution $f_i^\text{th, angled, ent}$ tested in this section. Analogous to equation (\ref{eqn: 1D Entropic Solution Truncation}),  we define the order $N$ truncation to be $[f_i^\text{th, angled, ent}]_N = \sum_{n = 0}^N P_n(\tau) (\partial_t + v_{i\alpha}\partial_\alpha)^n \feqi (\rho, \mathbf{u})$, where we replace $f_i^\text{eq}$ is by $f_i^\text{eq, ent}$. Because we observed, as shown below, that the formal analytical solution isn't consistent with the Couette flow, we focus here at the reason for this. In particular, the ensuing analysis predominantly looks at the induced macroscopic velocity profile of the proposed solution, whose correctness is a necessary condition for an analytical solution.

The truncation of order 1 of the formal solution preserves mass. In the following derivation of this fact and the next, we will keep the local equilibrium distribution and other variables arbitrary. After inserting chain rule, the order $1$ truncation of equation (\ref{eqn: simplified fi completely analytical general solution with polynomials}) is

\begin{align}
    (f_i)_1(\mathbf{x}, t) 
    =& f_i^\text{eq}(\rho, \mathbf{u}) - \tau (\partial_t \rho + v_{i\alpha} \partial_\alpha \rho) \frac{\partial f_i^\text{eq}}{\partial \rho} \nonumber\\
    &-\tau(\partial_t u_\alpha + v_{i\beta}\partial_\beta u_\alpha) \frac{\partial f_i^\text{eq}}{\partial u_\alpha},
     \label{eqn: 1st order truncation of general solution}
\end{align}
with Einstein notation used. To take the zeroth order velocity moment, we may bring the sum into the corresponding derivatives to get $\frac{\sum_i f_i^\text{eq}}{\partial \rho}$ and $\frac{\sum_i f_i^\text{eq} v_{i\beta}}{\partial u_\alpha}$. Upon substituting the zeroth and first moments of $f_i^\text{eq}$, we obtain
\begin{align}
    \sum_i (f_i)_1(\mathbf{x}, t) =&\rho - \tau \partial_t \rho - \tau u_\alpha \partial_\alpha \rho - \tau (\partial_\beta u_\alpha) \frac{\partial (\rho u_\beta)}{\partial u_\alpha} \nonumber \\
    =& \rho - \tau[\partial_t \rho + \partial_\alpha (\rho u_\alpha)] \nonumber \\
    =& \rho.
    \label{eqn: Mass moment of first order truncation of 2D entropic}
\end{align}
The last equality follows by application of the continuity equation of the inclined Couette flow. In summary, the order $1$ truncation conserves mass for any choice of local equilibrium distribution $f_i^\text{eq}$.

The order $1$ truncation does not preserve momentum. To analyze the first order velocity moment, we first define the quantity 
\begin{equation}
    P_{\alpha \beta} = \sum_i f_i^\text{eq} v_{i\alpha}v_{i\beta}.
    \label{eqn: Palphabeta discrete quantity}
\end{equation}
To take the first moment of equation (\ref{eqn: 1st order truncation of general solution}), we may bring the sum into the necessary derivatives like last time. Upon substituting the correct moments of $f_i^\text{eq}$ along with the definition of $P_{\alpha \beta}$, the result is
\begin{align}
    &\sum_i f_i(\mathbf{x}, t) v_{i\alpha} \nonumber\\
    =& \rho u_\alpha - \tau u_\alpha (\partial_t \rho) - \tau (\partial_\beta \rho) \frac{\partial P_{\alpha \beta}}{\partial \rho}\nonumber - \tau (\partial_t u_\beta) \rho \delta_{\alpha \beta} \nonumber\\
    &- \tau (\partial_\beta u_\gamma) \frac{\partial P_{\alpha \beta}}{\partial u_\gamma} \nonumber\\
    =& \rho u_\alpha - \tau \left[\partial_t (\rho u_\alpha) + (\partial_\beta \rho)\frac{\partial P_{\alpha \beta}}{\partial \rho} + (\partial_\beta u_\gamma)\frac{\partial P_{\alpha \beta}}{\partial u_\gamma} \right], \label{eqn: 2nd moment of 1st order truncation of general solution}
\end{align}
which holds for the order $1$ truncation of equation (\ref{eqn: simplified fi completely analytical general solution with polynomials}) using an arbitrary local equilibrium distribution. Going any further in the analysis requires implementing specific equilibrium distributions and flow profiles.

We computed $P_{\alpha \beta}$ for the entropic distribution using a computer algebra system and then verified it numerically on a number of different cases. It turns out to be
\begin{align}
    P_{xx} &=\frac{\rho}{3}(2\sqrt{1+3u_x^2}-1),\\
    P_{xy} &=\rho u_x u_y,\\
    P_{yy} &=\frac{\rho}{3}(2\sqrt{1+3u_y^2}-1).
\end{align}
The concise nature of this $P_{\alpha \beta}$ and the simplicity of the inclined Couette flow allows simplification of equation (\ref{eqn: 2nd moment of 1st order truncation of general solution}). The time derivative of momentum and spatial derivative of density both vanish, leaving
\begin{equation}
    \sum_{i} f_i(x,y) v_{i \alpha} = \rho u_\alpha - \tau (\partial_\beta u_\gamma)\frac{\partial P_{\alpha \beta}}{\partial u_\gamma}.
\end{equation}
If the Couette flow is aligned to the lattice ($\theta = 0$), then all but one spatial derivative of fluid velocity vanish, giving
\begin{align}
    \sum_{i} f_i(y) v_{ix} &= \rho u_x - \tau \dot{\gamma} \frac{\partial}{\partial u_x}(P_{xy}) = \rho u_x,\\
    \sum_{i} f_i(y)v_{iy} &= \rho u_y - \tau \dot{\gamma}\frac{\partial}{\partial u_x}(P_{yy}) = \rho u_y,
\end{align}
in which we observe momentum conservation. We expect this since we achieved macroscopic flow agreement using in the order $1$ entropic distribution aligned case, as mentioned in section \ref{sec: numerical examination, aligned entropic case}.

However, in the general angled case, we must contend with cross-terms depending on the angle, and equation (\ref{eqn: 2nd moment of 1st order truncation of general solution}) becomes the following:
\begin{widetext}
\begin{align}
    \sum_i f_i(x,y)v_{ix} &= \rho u_x(x,y) -\rho \tau \left[-\frac{2u_x(x,y)}{\sqrt{1+3u_x(x,y)^2}}\dot{\gamma} \sin(\theta) \cos(\theta) + u_y(x,y) \dot{\gamma} \cos^2(\theta) + u_x(x,y) \dot{\gamma} \sin(\theta) \cos(\theta) \right], \label{eqn: 2nd moment of 1st order truncation x}\\
    \sum_i f_i(x,y) v_{iy} &= \rho u_y(x,y) - \rho \tau \left[ -u_y(x,y)\dot{\gamma} \sin(\theta) \cos(\theta) - u_x(x,y)\dot{\gamma}\sin^2(\theta) + \frac{2u_y(x,y)}{\sqrt{1+3u_y(x,y)^2}} \dot{\gamma}\sin(\theta) \cos(\theta)\right]. \label{eqn: 2nd moment of 1st order truncation y}
\end{align}
\end{widetext}

\begin{figure}
    \centering
    \includegraphics[width = \linewidth]{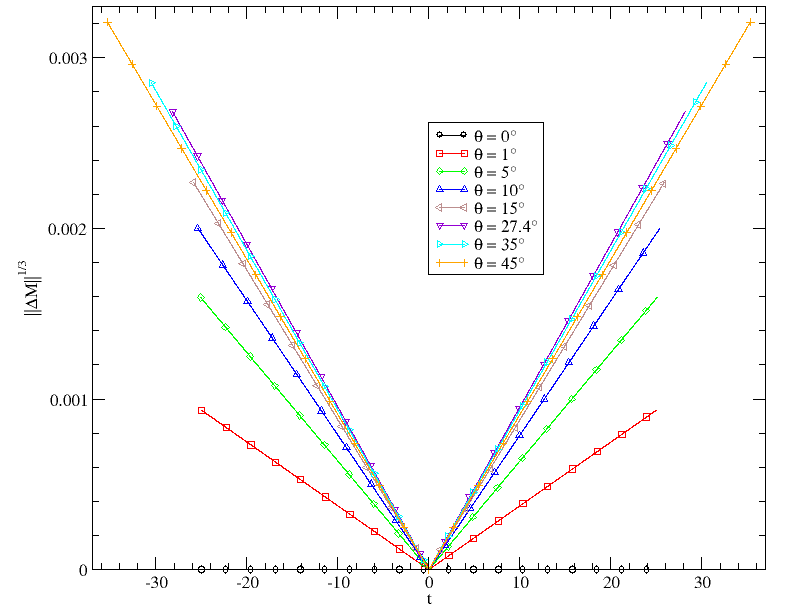}
    \caption{Example of wedge nature of $||\Delta \textbf{M}(\theta, x[t], y[t])||^{1/3}$ against the parameter $t$ for various angles $\theta$. The simulation parameters are $\dot{\gamma} = 0.001$, $H = 25$, $W = 25$. The specific values of $t$ are integer multiples of the step-size of $\frac{90}{4999} \approx 0.018$, within the appropriate range.}
    \label{fig: Example of wedge shape}
\end{figure}

In general, equations (\ref{eqn: 2nd moment of 1st order truncation x}) and (\ref{eqn: 2nd moment of 1st order truncation y}) are not $\rho u_x, \rho u_y$, respectively. To see this, consider the vector 
\begin{equation}
    \Delta\textbf{M}(\theta, x,y) =
    \begin{pmatrix}
        \frac{1}{\rho}[\partial_\beta u_\gamma] \frac{\partial P_{x\beta}}{\partial u_\gamma} \\
        \frac{1}{\rho}[\partial_\beta u_\gamma] \frac{\partial P_{y\beta}}{\partial u_\gamma}
    \end{pmatrix},
\end{equation}
whose components are explicitly given by the bracketed parts in equations (\ref{eqn: 2nd moment of 1st order truncation x}) and (\ref{eqn: 2nd moment of 1st order truncation y}). We can use the norm of $\Delta M$ measured in a direction orthogonal to the flow direction (\textit{i.e.} following the dashed line in Figure \ref{fig: angled couette flow diagram}) to examine how this error depends on the angle $\theta$. Parameterizing this line as $(x(t), y(t))$, with $t=0$ corresponding to the origin, we obtain all $t$-values such that the line segment lies within the simulation domain $[-25, 25]\times[-25,25]\subseteq \mathbb{R}^2$. The reason for taking our simulation domain to be this larger size as compared to the sizes in section \ref{sec: Numerical Examination of Aligned Shear Flow} is that we are more interested in seeing boundary effects. Furthermore, taking the cube root, the quantity $||\Delta \textbf{M}(\theta, x[t], y[t])||^{1/3}$ becomes a particularly simple wedge-shape. Figure \ref{fig: Example of wedge shape} provides some examples of this wedge shape against the associated parameter $t$ for various $\theta$ angles. Notice that as $\theta$ changes, the range of the corresponding $t$-values changes, which we expect since the length of a chord of a square, passing through the center, changes according to the angle of the chord.

\begin{figure}
    \centering
    \includegraphics[width=\linewidth]{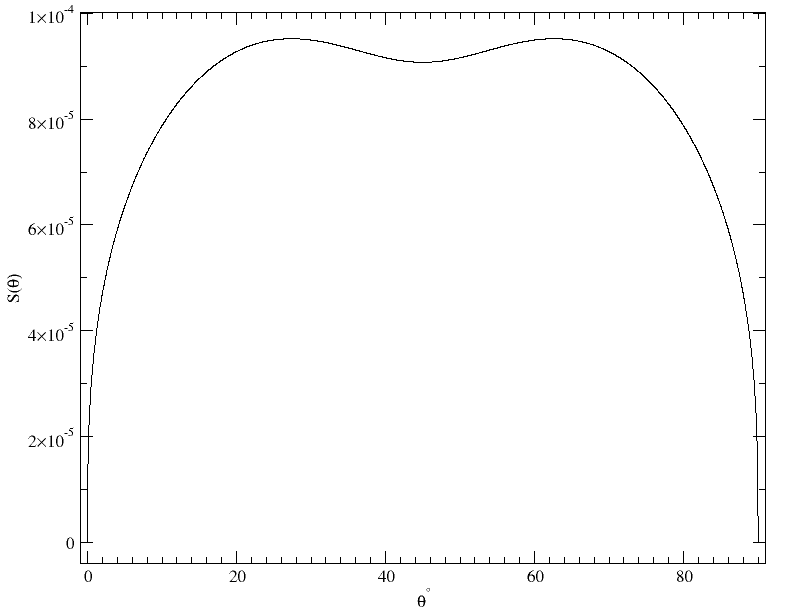}
    \caption{Plot of $S(\theta)$ for a fixed $\dot{\gamma} = 10^{-3}$ and a continuum of $\theta$ values between $0^\circ$ and $90^\circ$. Global maxima occur around $\theta = 27.4^\circ$ and $\theta = 62.6^\circ$, and there is a relative minimum at $\theta = 45^\circ$.}
    \label{fig: Wedge Graph, shows inacurracy of 2nd moment of 1st order solution}
\end{figure}

The slope of either side of the symmetrical wedges provides a measurement of deviation of the $1$st-velocity moments of the order $1$ truncation depending on angle $\theta$ and shear rate $\dot{\gamma}$. For a fixed $\dot{\gamma}$, the slope of the $t > 0$ side of the wedge shapes is 
\begin{equation}
    S(\theta) := \frac{||\Delta \textbf{M}(\theta, x[s], y[s])||^{1/3}}{s}
    \label{eqn: Slope of error measure of 2nd moment deviation},
\end{equation}
where $s$ is the parameter step-size.The results are essentially independent of $s$ but for this analysis we choose it to be $s=\frac{90}{4999}$. Figure \ref{fig: Wedge Graph, shows inacurracy of 2nd moment of 1st order solution} shows a plot of this error-slope for $\dot{\gamma}=0.001$. We see that the error does go to zero for flows aligned with the lattice but become larger for non-aligned flows. Perhaps surprisingly the maximum error is not found for $45^\circ$ but is rather closer to $27.4^\circ$. Note, however, that the deviation is rather small, the normalized error remaining less than 0.01 \% an error that would not be detectable in a vector plot for the velocity. 

The consequence of the mismatch of the momentum between the analytical solution $f^{th}$ and imposed momentum $\rho u$ for the order $1$ truncation is that the first order Chapman-Enskog analysis of equation (\ref{eqn: simplified fi completely analytical general solution with polynomials}) with $f_i^\text{eq}$ replaced by $f_i^\text{eq, ent}$ would not recover the Navier-Stokes equations but rather a slight modification to them. Thus one would expect that our entropic solution would be correct for a flow profile that satisfies these modified Navier-Stokes equations. One specific feature of the solution obtained by the entropic lattice Boltzmann method is that the density will not be constant, but rather be larger in areas with higher velocities. But for practical simulations those variations remain minor.

\begin{figure}
    \begin{subfigure}{0.82\linewidth}
    \includegraphics[width=\linewidth]{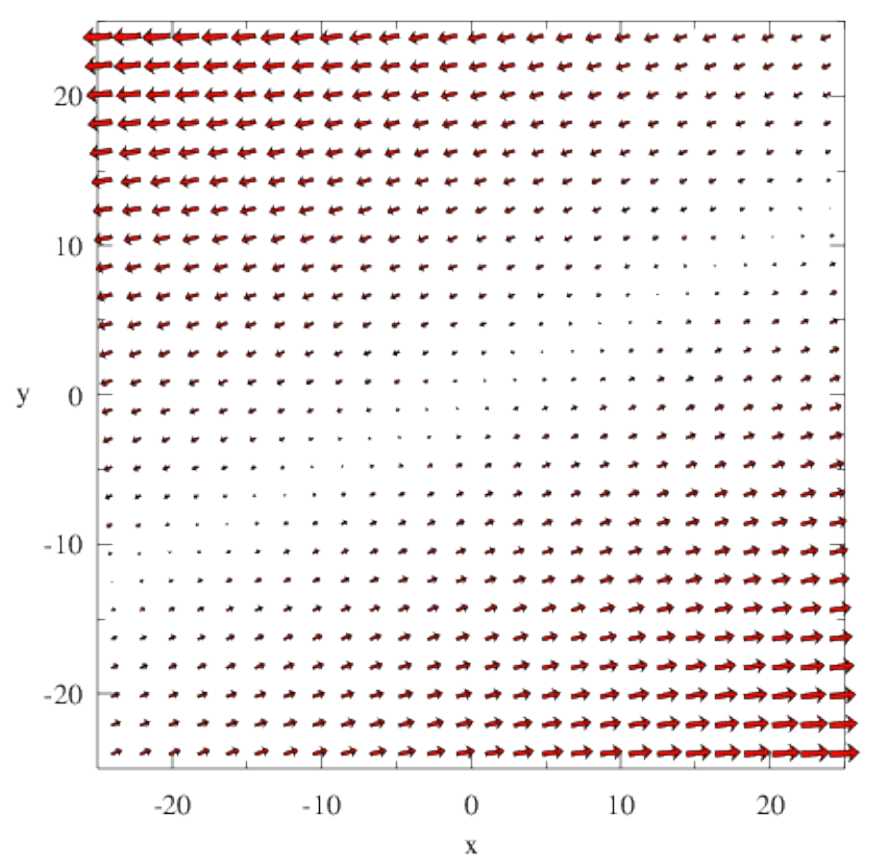}
        \caption{$\Delta \mathbf{E}(x,y)$ single step}
        \label{fig: Entropic Equilibrium One Step Velocity Profile Deviation}
    \end{subfigure}
    \begin{subfigure}{0.82\linewidth}
        \includegraphics[width=\linewidth]{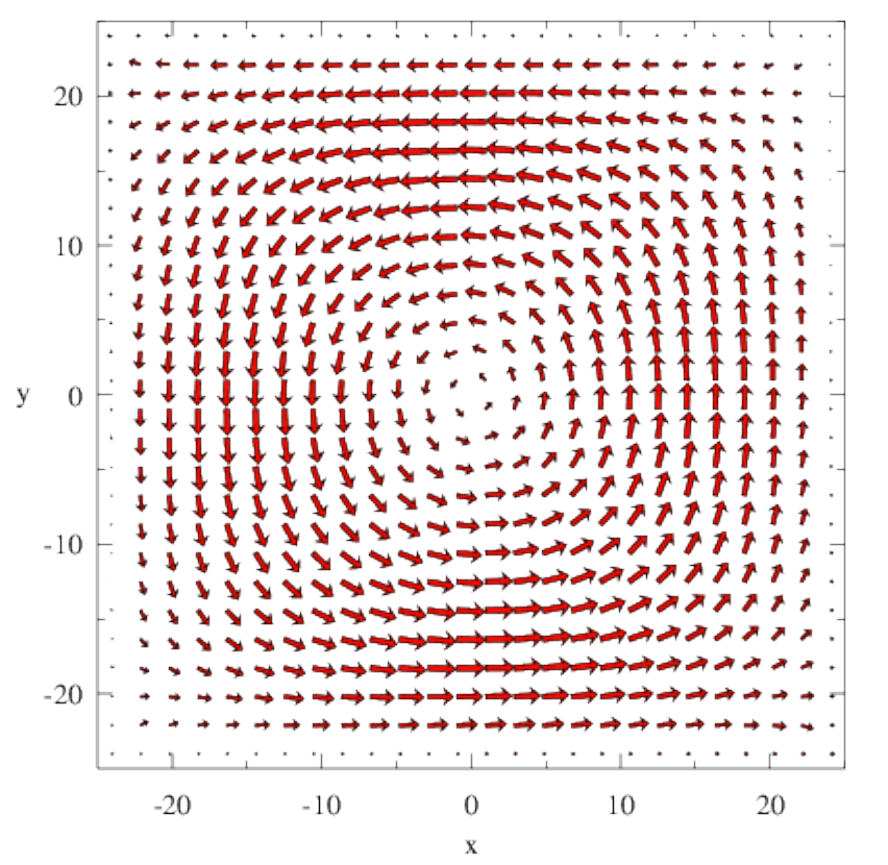}
        \caption{$\Delta \mathbf{E}(x,y)$ steady-state}
        \label{fig: Entropic Equilibrium STEADY STATE Velocity Profile Deviation}
    \end{subfigure}
    \caption{Incompatibility of Couette flow with entropic lattice Boltzmann. We define the relative error $\Delta \mathbf{E}$ by equation (\ref{eqn: relative uvecdiff error}). (a) Simulation parameters are $\theta = 27.4^\circ$, $\dot{\gamma} = 10^{-3}$, $\tau = 1$, dimension size $H = 25$, $W = 25$. The maximum vector length in this graph is $1.42 \cdot 10^{-6}$. We also omit every other vector for visibility purposes (b) Steady-state simulation results for the injection boundary condition and initialization of equation (\ref{eqn: Hazi Method}) (exact procedure mentioned at the start of section \ref{sec: Inclined Couette flow}. Same parameters as those in Figure \ref{fig: Entropic Equilibrium One Step Velocity Profile Deviation}. Maximum vector length in this graph is $6.57 \cdot 10^{-5}$, and we skip plotting every other vector.}
\end{figure}

In the special case of $\tau = 1$ for a steady-state flow, there is a trivial way of obtaining an analytical solution when we know the conserved fields. After collision, the $f_i$ will be equal to their equilibrium value, so we have
\begin{equation}
    f^\text{th, H\'azi}_i(\mathbf{x}) = f_i^\text{eq, ent}(\rho[\mathbf{x}-\mathbf{v}_i],\mathbf{u}[\mathbf{x}-\mathbf{v}_i]).
    \label{eqn: Hazi Method}
\end{equation}
This is the method, mentioned by H\'azi \cite{Hazi2003}, to check the compatibility of inclined Couette flow with entropic lattice Boltzmann. 

Once we set equation (\ref{eqn: Hazi Method}), then we may compute the induced velocity profile $\mathbf{u}(x,y)$ and compare it with that of the inclined Couette flow profile $\mathbf{u}^\text{th}(x,y)$. We define relative error as
\begin{align}
    &\Delta \mathbf{E}(x,y) \nonumber\\
    =& \frac{\mathbf{u}(x,y) - \mathbf{u}^\text{th}(x,y)}{\text{max}\{\mathbf{u}^\text{th}(x,y) : (x,y) \in [-25, 25] \times [-25,25]\}}.
    \label{eqn: relative uvecdiff error}
\end{align}
Figure \ref{fig: Entropic Equilibrium One Step Velocity Profile Deviation} shows $\Delta \mathbf{E}(x,y)$ for a specific configuration. Other configurations, except those for which $\theta$ is a multiple of $90^\circ$, have similar disagreement. We also note that the density profiles also disagree. We thus conclude that equation (\ref{eqn: Hazi Method}) is not an analytical solution. The non-vanishing of the higher moments mentioned in \cite{blommel2018integer} is perhaps manifesting itself here. This implies that the assumption that inclined Couette flow, oriented to an arbitrary angle, corresponds to solution of the entropic lattice Boltzmann method is false.

We may also run a simulation to steady state with equation (\ref{eqn: Hazi Method}) as the boundary condition to check how the magnitude of the error evolves. Figure \ref{fig: Entropic Equilibrium STEADY STATE Velocity Profile Deviation} indicates that the disagreement does not grow unboundedly and that the relative error is quite small. Evidently the angle of the Couette flow introduces secondary vortices dependent on the boundary conditions that create an asymmetry in the flow.

\section{Outlook}
In this paper, we presented a method for finding the lattice Boltzmann densities $f_i(\mathbf{x},t)$ for situations in which we know the macroscopic flow fields $\rho(\mathbf{x},t)$ and $\mathbf{u}(\mathbf{x},t)$. This approach recovered known analytical solutions as well as derive new ones. It also uncovered the incompatibility of these flow solutions with some lattice Boltzmann methods.

While we only applied this approach to a very simple problem, the problem of Couette flow, we demonstrated several interesting features. The solutions may be either in closed form or in the form of an infinite series. If the solution is finite, it may be relatively straight forward to prove that the solution found is actually the correct solution by inserting it into the lattice Boltzmann equation.

We may also use our method to show that certain flows are incompatible with a particular lattice Boltzmann approach. Here we showed that the entropic lattice Boltzmann method is not compatible with the Couette flow for flows that are not aligned with the lattice.

Such insights can be trivially used to devise boundary conditions to recover known flow profiles. A more interesting question is to what degree such boundary conditions vary for different analytically known flows, e.g. whether such inclined boundary conditions depend on the nature of the flow near to the wall.

Lastly, we believe that our approach will be very helpful for the inverse problem: if one knows the $f_i(\mathbf{x}, t)$, can one discover the underlying lattice Boltzmann method? This problem is of particular interest to us, for the Molecular Dynamics Lattice Gas (MDLG) approach \cite{parsa2017lattice} allows us to find lattice gas representations of underlying Molecular Dynamics simulations. These lattice gases can be averaged to give the lattice Boltzmann densities $f_i(\mathbf{x}, t)$ either directly from the MD simulation or from an underlying fundamental theory of particle displacements \cite{pachalieva2021connecting}. One can then use our approach to directly test whether any candidate lattice Boltzmann approach is consistent with the MDLG solution. If a more general set of lattice Boltzmann approaches can be parametrized, then one may use a learning procedure to optimize the match between the two.

\appendix
\section{Explicit Solutions for Couette Flow}
\label{appenddix: big equations}
In this appendix we spell out the explicit terms contained in equation (\ref{eqn: simplified fi completely analytical general solution with polynomials}). First, inserting the chain rule and multinomial expansions into equation (\ref{eqn: simplified fi completely analytical general solution with polynomials}) for an arbitrary flow gives (\ref{eqn: huge chain rule with polynomial}) as the formal analytical solution for any flow. 
\begin{widetext}
\begin{align}
    &f_i(\mathbf{x},t) \nonumber \\
    &= \sum_{n=0}^\infty \Delta t^n P_n(\tau) \sum_{\substack{i_1+i_2+i_3 = n, i_h \geq 0 \\ j_1+j_2+j_3 = i_3, j_h \geq 0 \\ k_1+k_2+k_3 = i_2, k_h \geq 0 \\ \ell_1+\ell_2+\ell_3 = i_1, \ell_h \geq 0}} (v_i^x)^{i_2} (v_i^y)^{i_3} {n \choose i_1, i_2, i_3} {i_1 \choose \ell_1, \ell_2, \ell_3}  {i_2 \choose k_1, k_2, k_3} {i_3 \choose j_1, j_2, j_3} \nonumber\\
    &\left( \partial_t \rho \right)^{\ell_1}
    \left(\partial_x \rho \right)^{k_1} \left( \partial_y \rho\right)^{j_1} \left( \partial_t u_x\right)^{\ell_2}  \left( \partial_t u_y \right)^{\ell_3} \left( \partial_x u_x \right)^{k_2} \left( \partial_x u_y\right)^{k_3} \left( \partial_y u_x \right)^{j_2} \left( \partial_y u_y \right)^{j_3} \frac{\partial^n f_i^\mathrm{eq}(\rho, \mathbf{u})}{\partial \rho^{j_1 + k_1 + \ell_1} \partial {u_x}^{j_2 + k_2 + \ell_2} \partial u_y^{j_3 + k_3 + \ell_3}},
    \label{eqn: huge chain rule with polynomial}
\end{align}
\end{widetext}
where 
\begin{equation}
    {n \choose i_1, i_2, i_3} := \frac{n!}{i_1!i_2!i_3!}.
\end{equation}
The application of this formula to the case of the quadratic local equilibrium distribution and an aligned Couette flow is straight forward and the result is given in equation (\ref{eqn: quadratic couette flat analytical solution}).

For the entropic local equilibrium distribution one may still simplify formula (\ref{eqn: huge chain rule with polynomial}). We first define
\begin{equation}
    f_y = 1+3v_{iy}u_y + (3v_{iy}^2 - 1)\left(\sqrt{1 + 3u_y^2} - 1\right).
\end{equation}
Recall that lattice units are set to $1$. Note that in the case of Couette flow of equation (\ref{eqn: Flat Couette Velocity Profile}) we have $u_y=0$ so we simply get $f_y = 1$. Upon using the simplifying features of the flow, equation (\ref{eqn: huge chain rule with polynomial}) becomes
\begin{equation}
    f_i^\text{th, ent} = \sum_{n = 0}^\infty f_i^\text{th, ent, n},
\label{eqn:FlatCouetteEntropicAnalSol}
\end{equation}
where 
\begin{widetext}
\begin{equation}
f_i^\text{th, ent, n} = 
    \begin{cases}
        f_i^\text{eq, ent}(\rho, u_x, u_y), & n = 0, \\
        - w_i \rho f_y \tau v_{iy} \dot{\gamma} \left[ 3v_{ix} + 3(3v_{ix}^2 - 1) \frac{u_x}{\sqrt{1+3u_x^2}} \right], & n = 1,\\
        w_i \rho f_y P_n(\tau) v_{iy}^n \dot{\gamma}^n (3v_{ix}^2 - 1) \frac{n! \sqrt{1+3u_x^2}}{2^n} \sum_{k \in \mathbb{Z}: \;n/2 \leq k \leq n} \frac{(-1)^{k-1} 6^k (2k-3)!!}{k!} {k \choose n-k} \frac{(u_x)^{2k-n}}{(1+3 u_x^2)^k}, & n \geq 2.
    \end{cases}
    \label{eqn: fithentN}
\end{equation}
\end{widetext}
Here the double factorial is defined as $n!! = 1$ for $n \leq 0$ and 
\begin{equation}
    n!! := \prod_{d=0}^{{\lfloor \frac{n-1}{2}\rfloor}} (n-2d)
\end{equation}
otherwise. Additionally, when $b > a$, we take ${a \choose b} = 0$, and we also put $0^0 = 1$. We provide a derivation of equation (\ref{eqn: fithentN}) in \cite{JordanSupplementalMaterial}, specifically the document titled, "Derivation\_of\_Equation\_A5."

For the inclined Couette flow with the quadratic distribution, equation (\ref{eqn: huge chain rule with polynomial}) becomes
\begin{equation}
    f_i^{\text{th, angled, pol}} = f_i^\text{eq, pol} + f_i^1 + f_i^2,
    \label{eqn: quadratic 2D Written Python}
\end{equation} where $f_i^\text{eq, pol}$ is the quadratic equilibrium distribution and $f_i^1, f_i^2$ are the correction terms given by 
\begin{widetext}
    \begin{align}
        f_i^1 =& w_i \rho \dot{\gamma} \tau \left( \frac{3}{2} \sin(2 \theta)[v_{ix} \{v_{ix} - u_x + 3v_{ix} \mathbf{v}_i \cdot \mathbf{u}\} - v_{iy} \left\{v_{iy} - u_y + 3v_{iy} \mathbf{v}_i \cdot \mathbf{u}\right\}] \right.\nonumber\\
        &+ 3 \sin(\theta)^2v_{ix} [v_{iy} - u_y + 3v_{iy} \mathbf{v}_i \cdot \mathbf{u}] \nonumber\\
        &-\left.3\cos(\theta)^2 v_{iy} [v_{ix} - u_x + 3v_{ix} \mathbf{v}_i \cdot \mathbf{u}] \right)
        \label{eqn: quadratic 2D Written Python first order correction}
    \end{align}
    and
    \begin{align}
        f_i^2 =& 3 \rho w_i \dot{\gamma}^2 \tau (2\tau-1) \left( 
        \sin^2(\theta)\cos^2(\theta)\left[v_{ix}^4 + v_{iy}^4 - 6\{v_{ix}v_{iy}\}^2\right] \right. \nonumber \\
        &+ \sin^3(\theta) \cos(\theta)[3v_{ix}v_{iy}\{v_{ix}^2-v_{iy}^2\}+v_{ix}v_{iy}] \nonumber\\
        &+\sin(\theta)\cos(\theta)^3[3v_{ix}v_{iy}\{v_{iy}^2-v_{ix}^2\}+v_{ix}v_{iy}]\nonumber\\
        &+\frac{\sin^4(\theta)}{2} \left[ 3\{v_{ix}v_{iy}\}^2 - v_{ix}^2\right] \nonumber\\
        &\left. +\frac{\cos^4(\theta)}{2}\left[ 3\{v_{ix}v_{iy}\}^2 - v_{iy}^2 \right] \right).
        \label{eqn: quadratic 2D Written Python second order correction}
    \end{align}
\end{widetext}
which, as we showed in the main text, is identical to writing the aligned solution and then rotating the velocity field.

\bibliography{AW,bib}

\end{document}


The following document provides a derivation of equation (A5) from the main text. 

\textbf{Bell Polynomials.} Consider the special case of $B_{n,k}(x,1,0,0,...,0)$ where $(x,1,...,0)$ is $n-k+1$ terms long. Let $$n-k+1 \geq 2.$$ Note that by definition of triangular array, $n \geq k$.

In the case $$2k-n \geq 0,$$ we need $j_1+j_2 = k$ and $j_1 + 2j_2 = n$. This gives $j_1 = 2k-n \geq 0$ and $j_2 = n-k \geq 0$. So in the sum in the definition of Bell polynomials, we have only one term. We get
\begin{equation}
    B_{n,k}(x,1,0,...,0) = \frac{n! x^{2k-n}}{2^{n-k}(2k-n)!(n-k)!},
    \label{eqn: Bell x,1,0,...}
\end{equation}
using the convention $0^0 = 1$ if necessary (matches C code convention for $\text{pow}(0,0)$). This expression for $B_{n,k}$ matches the Mathematica BellY function after a large number of tests. Property of Bell polynomial states 
\begin{equation}
    B_{n,k}(6x,6,0,...,0) = 6^k B_{n,k}(x,1,0,...,0).
    \label{eqn: Bell 6x, 6, 0, ...}
\end{equation}
This matches Mathematica as well.

In the case $$2k-n < 0,$$ we get an empty sum and so $$B_{n,k}(x,1,0,...,0) = 0,$$ which matches Mathematica.

\textbf{The Derivatives.} We must compute $$\frac{d^n}{d(u_x)^n} \left\{1-A_x+3v_{ix}u_x + A_x\sqrt{1+3u_x^2} \right\}.$$ We get
\begin{equation}
    \frac{d^n}{d(u_x)^n} \left\{1-A_x+3v_{ix}u_x + A_x\sqrt{1+3u_x^2} \right\} =
    \begin{cases}
    1-A_x+3v_{ix}u_x + A_x\sqrt{1+3u_x^2} & n = 0,\\
    3v_{ix} + A_x \frac{3u_x}{\sqrt{1+3u_x^2}} & n = 1,\\
    A_x\frac{d^n}{d(u_x)^n} \sqrt{1+3u_x^2} & n \geq 2.
    \end{cases}
    \label{eqn: Unexpanded nth deriv of fx}
\end{equation}

Set $f(x) = \sqrt{x}$ and $g(x) = 1+3x^2$. Then we see we must compute $$\frac{d^n}{dx^n} f(g(x)) = \sum_{k=0}^n f^{(k)}(g(x)) \cdot B_{n,k}(g'(x), g''(x), ..., g^{(n-k+1)}(x))$$ according to Wikipedia page on Fa\`a di Bruno's formula. This sum is 
\begin{align*}
    \frac{d^n}{dx^n} f(g(x)) &= f(g(x))B_{n,0}(6x,6,0,...,0) \\
    &+f'(g(x))B_{n,1}(6x,6,0,...,0)\\
    &+...\\
    &+f^{(n-2)}(g(x))B_{n,n-2}(6x, 6, 0)\\
    &+ f^{(n-1)}(g(x))B_{n,n-1}(6x,6)\\
    &+ f^{(n)}(g(x))B_{n,n}(6x),
\end{align*}
where the $B_{n,0}$ has $n+1$ inputs, $B_{n+1}$ has $n$ inputs, and so on.

By above, $B_{n,0}, ...$ are all $0$ until $2k - n \geq 0$. Also, $n \geq 2$ implies that we may use equations (\ref{eqn: Bell x,1,0,...}) and (\ref{eqn: Bell 6x, 6, 0, ...}) to compute the terms for which $2k -n \geq 0$ and $n-k+1 \geq 2$. The $n-k+1 \geq 2$ condition holds until the $B_{n,n}$ term. However, direct calculation shows that $B_{n,n}(6x) = 6^n x^n = \frac{6^n n! x^{2n-n}}{2^{n-n}(2n-n)!(n-n)!}$. 

Next check that
\begin{equation}
    f^{(k)}(x) = \frac{(-1)^{k-1} (2k-3)!! x^{-(2k-1)/2}}{2^k},
\end{equation}
for $k \geq 1$
where $j!! := j(j-2)(j-4)...$, the product terminating at $2$ or $1$, depending on whether $j$ is even or odd, respectively.

Now we combine the information of the last two paragraphs to compute the $n$th derivative for $n \geq 2$. Since $n \geq 2$ and we sum over $k$ such that $2k-n \geq 0$, we find $k \geq 1$, so we may use the equation of the last paragraph without restriction on $k$. Then 
\begin{align*}
    \frac{d^n}{dx^n} f(g(x)) &= \sum_{k: \;2k-n \geq 0, \;k \geq 0}^n \frac{(-1)^{k-1} (2k-3)!! (1+3x^2)^{-(2k-1)/2}}{2^k} \frac{6^k n! x^{2k-n}}{2^{n-k}(2k-n)!(n-k)!} \\
    &= \frac{n!\sqrt{1+3x^2}}{2^n}\sum_{k: \; 2k-n \geq 0, \; k \geq 0} \frac{(-1)^{k-1} 6^k (2k-3)!!}{(2k-n)!(n-k)!} \frac{x^{2k-n}}{(1+3x^2)^k}\\
    &= \frac{n!\sqrt{1+3x^2}}{2^n}\sum_{k: \; 2k-n \geq 0, \; k \geq 0} \frac{(-1)^{k-1} 6^k (2k-3)!!}{k!} {k \choose n-k}  \frac{x^{2k-n}}{(1+3x^2)^k}.
\end{align*}
Equation (\ref{eqn: Unexpanded nth deriv of fx}) becomes
\begin{equation}
    \frac{d^n}{d(u_x)^n} \left\{1-A_x+3v_{ix}u_x + A_x\sqrt{1+3u_x^2} \right\} =
    \begin{cases}
    1-A_x+3v_{ix}u_x + A_x\sqrt{1+3u_x^2} & n = 0,\\
    3v_{ix} + A_x \frac{3u_x}{\sqrt{1+3u_x^2}} & n = 1,\\
    A_x \frac{n!\sqrt{1+3x^2}}{2^n}\sum_{k: \; 2k-n \geq 0, \; k \geq 0} \frac{(-1)^{k-1} 6^k (2k-3)!!}{k!} {k \choose n-k}  \frac{u_x^{2k-n}}{(1+3u_x^2)^k} & n \geq 2.
    \end{cases}
    \label{eqn: nth deriv of fx wrt ux}
\end{equation}

\textbf{Putting it together.} Recall (in lattice units) that $$f_i^\text{eq, ent}(\rho, \mathbf{u}) = \rho w_i f_x(u_x) f_y(u_y)$$ where
\begin{align}
    f_x(u_x) &= 1-A_x + 3v_{ix}u_x + A_x\sqrt{1 + 3u_x^2},\\
    f_y(u_y) &= 1 - A_y + 3v_{iy}u_y + A_y \sqrt{1+3u_y^2},\\
    A_x &= 3v_{ix}^2 - 1,\\
    A_y &= 3v_{iy}^2 - 1.
\end{align}
We need to compute $$\frac{\partial^n f_i^\mathrm{eq, ent}(\rho, \mathbf{u})}{\partial \rho^{j_1 + k_1 + \ell_1} \partial {u_x}^{j_2 + k_2 + \ell_2} \partial u_y^{j_3 + k_3 + \ell_3}}.$$ In the Couette flow set up, this reduces to $$\frac{\partial^n f_i^\mathrm{eq, ent}(1, \mathbf{u})}{\partial {u_x}^{j_2 + k_2} \partial u_y^{j_3 + k_3}} = w_i \frac{d^{j_2 + k_2}f_x}{d(u_x)^{j_2+k_2}} \frac{d^{j_3+k_3}f_y}{d(u_y)^{j_3+k_3}}.$$ Equation (\ref{eqn: nth deriv of fx wrt ux}) along with its $f_y$ counterpart fully determines this formula. One may write this product explicitly via indicator functions, but it's not very insightful. The velocity derivatives of the flow profile are easy to compute. Thus, from here we easily get equation (A1) and therefore equation (A5).